\documentclass[a4paper,fleqn,usenatbib]{mnras}

\usepackage{newtxtext,newtxmath,amsmath}
\usepackage[T1]{fontenc}
\usepackage{ae,aecompl}
\usepackage{amsmath}
\usepackage{pbox}

\usepackage{graphicx}	% Including figure files
\usepackage{amsmath}	% Advanced maths commands
\usepackage{amssymb}	% Extra maths symbols
\usepackage{gensymb}
\usepackage{subfig}

\title[Radiation-hydrodynamics of thermally-driven disc winds in XRBs]{Radiation-hydrodynamic simulations of thermally-driven disc winds in X-ray binaries: A direct comparison to GRO~J1655-40}

\author[N. Higginbottom et. al]
{Nick Higginbottom,$^{1}$\thanks{E-mail: nick\_higginbottom@fastmail.fm}
Christian Knigge$^{1}$, Knox S. Long$^{2,3}$, 
James H. Matthews$^{4}$, \newauthor{Stuart A. Sim$^{5}$ and
Henrietta A. Hewitt$^{5}$.}
\\
$^{1}$School of Physics and Astronomy, University of Southampton, Highfield, Southampton, SO17 1BJ, UK\\
$^{2}$Space Telescope Science Institute, 3700 San Martin Drive, Baltimore, MD, 21218, USA\\
$^{3}$Eureka Scientific Inc., 2542 Delmar Avenue, Suite 100, Oakland, CA, 94602-3017, USA\\
$^{4}$University of Oxford, Astrophysics, Keble Road, Oxford OX1 3RH, UK\\
$^{5}$School of Mathematics and Physics, Queen's University Belfast, University Road, Belfast 
BT7 1NN, UK\\
}

\date{Accepted XXX. Received YYY; in original form ZZZ}

\pubyear{2018}

\hypersetup{draft}
\begin{document}
\label{firstpage}
\pagerange{\pageref{firstpage}--\pageref{lastpage}}
\maketitle

\begin{abstract}
Essentially all low-mass X-ray binaries (LMXBs) in the soft state appear to drive powerful equatorial disc winds. A simple mechanism for driving such outflows involves X-ray heating of the top of the disc atmosphere to the Compton temperature. Beyond the Compton radius, the thermal speed exceeds the escape velocity, and mass loss is inevitable. Here, we present the first coupled radiation-hydrodynamic simulation of such thermally-driven disc winds. The main advance over previous modelling efforts is that the frequency-dependent attenuation of the irradiating SED is taken into account. We can therefore relax the approximation that the wind is optically thin throughout which is unlikely to hold in the crucial acceleration zone of the flow. The main remaining limitations of our simulations are connected to our treatment of optically thick regions. Adopting parameters representative of the wind-driving LMXB GRO~J1655-40, our radiation-hydrodynamic model yields a mass-loss rate that is $\simeq5\times$ lower than that suggested by pure hydrodynamic, optically thin models. This outflow rate still represents more than twice the accretion rate and agrees well with the mass-loss rate inferred from Chandra/HETG observations of GRO~J1655-40 at a time when the system had a similar luminosity to that adopted in our simulations. The Fe~{\sc xxv} and Fe~{\sc xxvi} Lyman $\rm{\alpha}~$ absorption line profiles observed in this state are slightly stronger than those predicted by our simulations but the qualitative agreement between observed and simulated outflow properties means that thermal driving is a viable mechanism for powering the disc winds seen in soft-state LMXBs.
\end{abstract}

\begin{keywords}
Accretion discs -- hydrodynamics -- methods:numerical -- stars:winds -- X-rays:binaries
\end{keywords}

%%%%%%%%%%%%%%%%%%%%%%%%%%%%%%%%%%%%%%%%%%%%%%%%%%

%%%%%%%%%%%%%%%%% BODY OF PAPER %%%%%%%%%%%%%%%%%%

\section{Introduction}

Outflowing gas is seen in a vast range of accreting systems from cataclysmic variables \citep[e.g.][]{1982ApJ...260..716C,2011MNRAS.418L.129K} to active galactic nuclei (AGN) \citep[e.g.][]
{1982MNRAS.199..883B,1991ApJ...373...23W,2009A&ARv..17...47T}. These
outflows often have very high velocities and appear to transport a great deal of material from the accreting
object out to its environment. There are, broadly speaking, two classes of outflows: highly collimated fast polar jets \citep[e.g.][]{1997ARA&A..35..607Z,1999ARA&A..37..409M,2006ARA&A..44...49R} and weakly collimated disc winds \citep[e.g.][]{1994ApJ...434..446K,
1997ApJ...476..291K,2013AcPol..53..659D}. In general, the former are easier to observe because they can remain distinct over great distances. The latter tend to reveal themselves only via blue shifted absorption features, but have the potential for a greater impact on their surroundings, because they transport much more mass and momentum \citep{2012ARA&A..50..455F}.

This effect on their environment is one reason for the astrophysical importance of disc winds. Another is that the removal of mass and angular momentum from the disc might fundamentally change the accretion process itself \citep{1986ApJ...306...90S}. In addition, if such outflows are ubiquitous, then most observations of disc-accreting systems may contain signatures of these disc winds and need to be interpreted with this in mind.

Low-mass X-ray binaries (LMXBs) are excellent laboratories for studying accretion physics. First, they are much closer than the more dramatic AGN, and are therefore somewhat easier to observe. Second, and more importantly, they exhibit dramatic ``state changes'' over fairly short time-scales of days-weeks \citep{2010LNP...794...53B}. These shifts from a low-luminosity spectrally hard (low-hard) state to
a high-luminosity spectrally soft (high-soft) state are usually attributed to changes in the accretion disc \citep{2012MNRAS.422L..11P}. LMXBs spend most of their time in the low-hard state \citep{2010LNP...794...53B}, in which the inner part of the geometrically thin accretion disc is truncated far away from the central object. A hot, geometrically thick and optically thin accretion flow (aka ``corona'') takes over from the disc at smaller radii and produces most of the hard X-ray emission that dominates in this state. During a transition to the high-soft state the inner disc edge is thought to move inward  to smaller radii, thus increasing in luminosity and producing a softer overall spectrum. 

Clear observational evidence for strong disc winds has so far been seen in 8 LMXBs \citep{2016AN....337..368D}. In all cases, these systems were in the disc-dominated high-soft state. Moreover, 7 of the 8 sources are known to exhibit dips in their light curves - clear evidence that they are viewed edge on \citep{1987A&A...178..137F}. Taken together, this suggests that the outflows themselves are equatorial and associated with the accretion disc \citep{2012MNRAS.422L..11P}. The association between disc winds and high-soft states might also  point to a more fundamental role for these outflows in triggering state transitions. For example, if the wind transports enough mass away from the accretion disc, this might force the  system to change back to the low-hard state \citep{1986ApJ...306...90S}.

The driving mechanism for the disc winds seen in LMXBs is not clear. There are three main possibilities:  radiation driving \citep[e.g.][]{1980AJ.....85..329I,1985ApJ...294...96S,1993ApJ...409..372S,
2002ApJ...565..455P},
magneto-centrifugal driving \citep[e.g.][]{1982MNRAS.199..883B,1992ApJ...385..460E,2006Natur.441..953M,2016A&A...589A.119C} and thermal driving
\citep{1983ApJ...271...70B, 1996ApJ...461..767W, 2010ApJ...719..515L}. 

Radiation driving acting alone is probably the least promising of these candidates. Even in the high-soft state, disc luminosities tend to be below the Eddington limit, so continuum driving alone is usually not sufficient. Line driving is also problematic because the X-ray irradiated gas in these systems is too highly ionized \citep{2016AN....337..368D}. 

Magneto-centrifugal driving, where gas is loaded onto magnetic field lines threading the accretion disc and accelerated outwards along the lines like `beads-on-a-wire' offers a potential mechanism for winds that appear to arise very close in to the central object. This has been suggested as the driving mechanism for a very dense and fast outflow seen in GROJ1655-40 \citep[but also see \citealt{2006ApJ...652L.117N,2015MNRAS.451..475U,2016ApJ...823..159S} and Section \ref{section:GRO_params}]{1992ApJS...80..753S,2006Natur.441..953M,2008ApJ...680.1359M,2009ApJ...701..865K}. 

Thermal driving is a very simple and attractive mechanism for most of the disc winds seen in LMXBs. A thermally-driven outflow will arise naturally when the surface of the outer accretion disc is heated to the Compton temperature by X-rays produced in the inner disc and/or the corona. The location in the disc where the thermal velocity at the Compton temperature equals the local escape velocity is referred to as the Compton radius. Beyond this point, the thermal velocity in the upper disc atmosphere exceeds the escape velocity, so material will expand and escape in an out-flowing disc wind. 

In practice, the radius at which thermal driving becomes effective is about 10\% of the Compton radius \citep{1983ApJ...271...70B}. For very luminous sources, where the radiation pressure on electrons reduces the effective gravity significantly, the critical radius can be even smaller \citep{2002ApJ...565..455P}. Nevertheless, thermal winds preferentially arise at rather large distances from the central object. This is consistent with the observation that outflows are preferentially observed in long-period systems with large accretion discs \citep{2016AN....337..368D}.

Given that disc winds may play a crucial role in triggering state transitions, as well as their apparent ubiquity in disc-dominated states \citep{2012MNRAS.422L..11P}, it is important to 
understand the nature of these outflows. Moreover, since some thermally driven mass loss from the outer disc must occur in many systems, it is natural to explore this mechanism carefully first. Here, we therefore continue a project to numerically simulate thermally driven disc winds in LMXBs and  present a direct comparison of our most realistic simulation to observations.

A key assumption made in all simulations of thermally driven disc winds to date is that the outflow is completely optically thin throughout. This matters. The strength and character of these outflows depends strongly on the thermal stability of material in the disc atmosphere 
\citep[][hereafter HP15 and H17 respectively]{2015ApJ...807..107H,2017ApJ...836...42H}, and this in turn depends strongly on the shape of the irradiating SED \citep{2017MNRAS.467.4161D}. In an optically thin outflow irradiated by a central source, all parts of the wind see exactly the same irradiating spectrum, with a normalization set simply by geometric dilution. However, if the outflow is self-shielding, such approximations break down. To see this, consider a parcel of gas near the wind acceleration zone at some radius $R_1$ from the central source. If the irradiating SED reaching this parcel is strongly attenuated by absorption and/or scattering in the wind at radii $R < R_1$, the thermal state of this parcel can change dramatically, even to the point where it may no longer be significantly accelerated at all.

Given the sensitivity of thermally driven disc winds to self-shielding effects, it is clearly important to relax the $\tau << 1$ approximation. This is the goal of the present study. Of course, relaxing the ``no-self-shielding'' approximation turns a pure hydrodynamics problem into a radiation-hydrodynamics (RHD) one. As described in more detail below, we tackle this by coupling the well-known {\sc zeus} hydrodynamics code 
\citep[][extended by \citealt{2000ApJ...543..686P}]{1992ApJS...80..753S}, with our own state-of-the-art ionization and radiative transfer code {\sc python} \citep[][extended by \citealt{,2013MNRAS.436.1390H} and \citealt{2015MNRAS.450.3331M}]{2002ApJ...579..725L}. 

Our RHD simulations of thermally driven disc winds improve on existing work 
in two key areas. First, the attenuation of the irradiating SED by the opacity
in the outflow is fully taken into account. Second, the resulting spatially and
spectrally varying radiation field is used to estimate the thermal and ionization
state of the wind. As we shall see, self-shielding is {\em not} a second order 
effect. The local radiation field in key parts of the wind can differ drastically
-- by orders of magnitude at frequencies around the Hydrogen photo-ionization 
edge -- from the unattenuated X-ray spectrum produced by the central source. 
Perhaps unsurprisingly, this has a significant effect on the resulting outflow. As we 
shall see, the properties of the thermally driven wind seen in our RHD simulation 
are a surprisingly good match to those inferred observationally for the well-known wind-driving LMXB GRO~J1655-40 in its soft-intermediate state.

\section{GRO~J1655-40 as a benchmark}
\label{section:GRO_params}

We adopt system parameters designed to roughly describe the accreting black hole system GRO~J1655-40, a well-known wind-driving LMXB. This choice allows a direct comparison with previous work \citep[][HP15 and H17]{2010ApJ...719..515L} and provides us with high-quality observational data against which we can test our model. 

Specifically, we assume a central object mass of $7M_{\odot}$, a luminosity of $3.3\times10^{37}~\rm{ergs~s^{-1}(\simeq 0.04L_{Edd}})$ and the same bremsstrahlung-type SED used by HP15 and H17,  
\begin{equation}
L_{\nu}=K\nu \exp(-h\nu/k_BT_x).
\end{equation}
Given a typical accretion efficiency of 8.3\%, this luminosity implies a mass
accretion rate of $\dot{\rm{M}}_{\rm{acc}} = 4.4\times10^{17}~g~s^{-1}$. We use the same X-ray temperature as in H17, $T_x=5.6\times10^{7}K$, which corresponds to a Compton temperature of $T_{IC}=1.4\times10^{7}K$. 
 
It is crucial to note that these parameters are {\em not} appropriate for modelling the April 2005 {\em Chandra} observations of GRO~J1655-40 described by \citet[also see \citealt{2008ApJ...680.1359M} and \citealt{2009ApJ...701..865K}]{2006Natur.441..953M}, which were the basis for their claim of a magnetically driven outflow from the system. The original estimate of $L \simeq 0.04 L_{Edd}$ for this data set has been challenged by \citet[on the basis of the X-ray continuum]{2012MNRAS.422L..11P}, by \citet[on the basis of timing properties]{2015MNRAS.451..475U} and by \citet{2016ApJ...822...20N} and
\citet[on the basis of the OIR to X-ray SED]{2016ApJ...823..159S}. Several of these recent studies suggest that, at the time of these observations, the system was in an unusual ``hypersoft'' state characterized by $L > 0.1 L_{Edd}$, and indeed most likely $L \gtrsim L_{Edd}$ \citep{2015MNRAS.451..475U,2016ApJ...823..159S,2018MNRAS.473..838D}. This hypersoft state is very similar to that seen in the high-mass X-ray binary Cyg X-3, also thought to be accreting at or above the Eddington limit \citep{2015MNRAS.451..475U}.
It remains to be seen whether the need for a magnetic contribution to the wind-driving mechanism in the hypersoft state survives this revision. In any case, data obtained in the hypersoft state clearly cannot be used to test models of thermally driven disc winds in the normal soft state.

\begin{figure*}
\includegraphics[width=\textwidth]{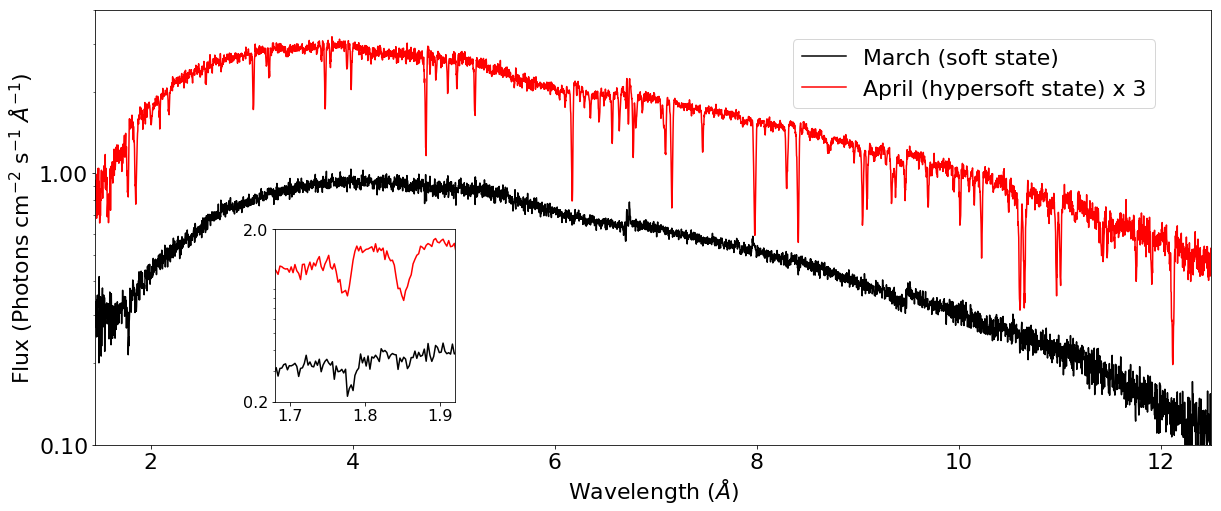}
\caption{A comparison of two {\em Chandra} HETG observations obtained during the 2005 outburst of GRO~J1655-40. \citet{2006Natur.441..953M,2008ApJ...680.1359M} and \citet{2009ApJ...701..865K} analysed the April (red) spectrum, which was taken during an unusual ``hypersoft'' state. By contrast,  the parameters adopted in our simulations are more representative of the soft-intermediate state during which the March (black) spectrum was obtained.}
\label{figure:states}
\end{figure*}

Fortunately, {\em Chandra} obtained another observation of GRO~J1655-40 just 20 days earlier, in March 2005 \citep{2008ApJ...680.1359M,2012ApJ...750...27N}. This observation took place close to the peak of the outburst and near the end of the hard-to-soft state transition. We will refer to the state of the system at this time as a ``soft-intermediate state''. \citet{2012MNRAS.422L..11P} estimate $L \simeq 0.04~L_{Edd}$ for this state from their modelling of the {\em Chandra+RXTE} X-ray continuum. As shown in Figure~\ref{figure:states}, the soft-intermediate state spectrum exhibits significantly fewer and weaker absorption lines than the hypersoft spectrum. In Section~\ref{section:comp_to_obs}, we will use this March 2005 spectrum as a testbed for our RHD simulations.

\section{Computational Method}
\label{section:method}
\subsection{Overview}

Our RHD simulations use an operator-splitting technique to couple {\sc zeus} to our Monte Carlo ionization and radiative transfer code {\sc python}.  
This technique permits a complete treatment of radiative transfer (RT) between
the source of radiation and each cell in the model. \textsc{python}
calculates the ionization state of the plasma using the local radiation field, which ensures that
energy deposition is calculated accurately, even for highly frequency dependent processes like 
photo-ionization. This represents a significant advance over previous efforts to model such disk winds, which have either assumed optically thin conditions or grey attenuation.

In the course of an RHD calculation, we continuously cycle between hydrodynamic (HD) steps carried
out by {\sc zeus} and ionization/radiative transfer (IRT) steps carried out by 
{\sc python}, with information about the state of the outflow being passed back 
and forth between the two codes. 
In each IRT step, {\sc python} takes the latest snapshot of the outflow (i.e.  the 
density, temperature and velocity field across the simulation grid) and injects a 
population of $N=10^{7}$ ``photons'' (strictly energy packets) that correctly 
represents the source SED.
The code then follows the passage of these photons through
the outflow, keeping track of the energy transfer between the radiation field and 
the plasma. This information is then used to estimate the ionization state of each
grid cell, as well as its heating and cooling rate. In practice, we actually carry
out two iterations of this process during each call to {\sc python}, in order to
ensure that our estimate of the ionization state is converged. Test calculations
confirm that two iterations are sufficient to achieve this.

Internally, the version of {\sc zeus} used here represents the heating and cooling
rates via modified versions of the approximate analytic formulae provided by \cite{1994ApJ...435..756B}. This is similar to the approach adopted by  \cite{2010ApJ...719..515L}, HP15 and  H17. Our versions of 
these formulae are given in Equations~ \ref{eq_comp_heat}-\ref{eq_brem_cool} below and discussed in the accompanying text. 

The standard normalization of these analytic approximations was calibrated for a specific incident SED. However, since our calculations account for attenuation effects, each cell in our simulation can see a different SED. We therefore re-calibrate each 
heating/cooling term in the analytic rates for each grid cell after each IRT step
by a multiplicative normalization factor - the $K$-terms in equations \ref{eq_comp_heat}-\ref{eq_brem_cool}. Since \textsc{python} is typically only called every
$t_{HD,max}$ simulation seconds (see below), we can then use these recalibrated formulae to 
continuously (if approximately) update the heating/cooling rates between calls 
to \textsc{python}. After each call to \textsc{python}, the heating and cooling
rates for each cell are imported into \textsc{zeus}
and these rates are compared to the heating and cooling rates used in the last hydrodynamic
cycle. The normalization factors are then adjusted to make the rates to be used in
the next cycle equal to those seen in the \textsc{python} simulation.

In order to avoid discontinuous changes in the heating and cooling rates, we apply a 
numerical damping factor during the recalibration after each IRT step. Since we expect -- and find -- that our converged simulation reaches a steady state, the exact way in which the simulation approaches this state is not critical. This numerical damping therefore helps with convergence, but does not affect the final results. We have carried out several tests to verify this, using different damping factors, for example. Our final damping factor for these runs is 0.9 - meaning that the calibration factors can only change by about 10\% after each IRT step.

Our simple implementation of RHD is computationally expensive. The computational 
effort is totally dominated by the IRT steps, each of which takes about $50\times$ 
longer than a single HD step, even with the IRT calculations running on 256 processors 
in parallel. It is therefore not (yet) feasible to follow every HD step with a new IRT
step. In order to overcome this problem, we inspect the properties of the outflow after
each HD step and call {\sc python} if more than 10\% of the cells have changed their 
density by more than 50\%, or if $t_{HD,max} = 1000$ s of simulation 
time have elapsed since then. Again, this treatment is reasonable, because our thermally driven disc 
winds tend to a steady state. We consider a run to be converged if the mass-loss rate is no longer 
changing significantly. We have verified that our value of  $t_{HD,max}$ is reasonable by carrying out test calculations with different $t_{HD,max}$ in a small simulation domain (meaning that the run can be converged in a reasonable time, but at the cost of no longer capturing the Mach surface). In these test runs, reducing $t_{HD,max}$ by an order of magnitude increases the mass-loss rate by about 30\%.

\subsection{Heating and cooling rates}
\label{section:heat_cool}

Our modified versions of the approximate analytic heating and cooling rate equations are as follows. First, we split up the Compton heating ($H_c$) and cooling ($\Lambda_c$) rates into two separate terms. Thus Compton heating is described by 
\begin{equation}
H_c=K_{H_c} 8.9\times10^{-36} \xi T_x ~\rm{(ergs~s^{-1}~cm^{3})},
\label{eq_comp_heat}
\end{equation}
while Compton cooling is modelled as 
\begin{equation}
\Lambda_c=K_{\Lambda_c} 8.9\times10^{-36}  \xi (4T)~\rm{(ergs~s^{-1}~cm^{3})}.
\label{eq_comp_cool}
\end{equation}
T is the electron temperature of the plasma, and the ionization parameter, $\xi$, is defined as
\begin{equation}
\xi=\frac{L_x}{n_Hr^2},
\label{eq:ip}
\end{equation}
where $L_x$ is the integrated luminosity above 13.6eV, $n_H$ is the Hydrogen number density, 
and r is the distance between a given cell and the origin.

Splitting the heating and cooling rates allows us to calibrate the two terms independently, via their respective $K$-factors. This is necessary because Compton heating and cooling depend differently on the illuminating SED. With our new treatment of attenuation, different cells will generally see different SEDs, so our on-the-fly re-calibration needs to treat these two terms separately.

The so-called line cooling term ($\Lambda_l$) includes contributions from both bound-bound
and free-bound interactions. Our new version of this term is
\begin{equation}
\Lambda_l=K_{\Lambda_l} \left( 1.0\times10^{-16} \frac{\exp{(-1.3\times10^{5}/T)}}{T\sqrt{\xi}}
+\mathcal{K}(T)\right)
\label{eq_line_cool}
\end{equation}
The units are the same as for Equation \ref{eq:ip} and 
\begin{align}
\label{eq_line_cool_cases}
\mathcal{K}(T) & = 
\begin{cases}
5.0 \times 10^{-27} \; T^{1/2} & \hspace*{0.5cm} ~~~~~~~~~~~T < 10^4~{\rm K} \\
1.0 \times 10^{-24} &\hspace*{0.5cm} 10^4 ~{\rm K} < T < 10^7~{\rm K} \\
1.5 \times 10^{-17}\; T^{-1} & \hspace*{0.5cm}~~~~~~~~~~~T > 10^7~{\rm K}. \\
\end{cases}
\end{align}

Blondin's (1994) original formulation corresponds to the intermediate temperature regime in Equation~\ref{eq_line_cool_cases}. However, in our simulations, we sometimes encounter dense regions near the mid-plane where the gas is no longer fully ionized. The low-temperature regime in Equation~\ref{eq_line_cool_cases} accounts for this onset of partial ionization, where the cooling rate starts to depend on the temperature-sensitive collisional ionization rate. Similarly the high-temperature regime in Equation~\ref{eq_line_cool_cases} accounts for the  roll-off in the recombination cooling rate when the gas becomes over-ionized. This actually has little impact, since the cooling rate at such high temperatures is dominated by Compton cooling. It should be noted that this modified formula on its own does not represent a better fit to line cooling rates over all
ionization parameters and temperatures. It is, however, a much better fit at low ionization
parameters and allows us to achieve excellent agreement with the IRT rates (in combination with the $K_{\Lambda_l}$ calibration factor).

The ``X-ray'' (photo ionization + Auger) heating rate ($H_x$) and bremsstrahlung
cooling rate ($\Lambda_b$) equations are unchanged from Blondin (1994), except for the normalization factors. We nevertheless include them here for completeness:
\begin{equation}
H_x=K_{H_x}1.5\times10^{-21} \sqrt{\xi/T}\left(1-\left(T/T_x\right)\right)~\rm{(ergs~s^{-1}~cm^{3})},
\end{equation}
and 
\begin{equation}
\Lambda_b=K_{\Lambda_b} 3.3\times10^{-27}\sqrt{T}~\rm{(ergs~s^{-1}~cm^{3})}.
\label{eq_brem_cool}
\end{equation}

All five heating and cooling rates are combined to give the total net heating rate 
\begin{equation}
\rho\mathcal{L}=n_H\left(n_eH_c+n_HH_x-n_e\Lambda_c-n_e\Lambda_l-n_e\Lambda_b\right).
\label{eq:total_heatcool}
\end{equation}
This equation is somewhat modified from the form given in H17, because here we
include the effect of partial ionization. Specifically, H17 used the multiplier of $n_e n_H$ for all heating rates, but the X-ray heating rate does not, in fact, depend on $n_e$. This is not significant 
for the optically thin HD simulations presented in H17, since $n_e = 1.21 n_H$ for a fully ionized plasma with solar abundances. However, as noted above, and discussed in detail in Section \ref{section:results}, in our RHD simulations a dense region forms towards the mid-plane. This region is not fully ionized due to photo-absorption, and so $n_e$ can no longer be approximated there as a simple multiple of $n_H$. We have therefore carried out a set of stand-alone \textsc{python} simulations covering the critical range of temperatures and fit the resulting electron to Hydrogen density ratio with the following piecewise fit:

\begin{align}
\label{equation:ne_cases}
\frac{n_e}{n_H} &= 
\begin{cases}
2.5\times10^{-52}~T^{12.3} & \hspace*{0.5cm} ~~~~~~~~~~~~~~T < 15000~{\rm K} \\
2.5\times10^{-3.8}~T^{0.86} &\hspace*{0.5cm} 15000 ~{\rm K} < T < 33000~{\rm K} \\
1.21 & \hspace*{0.5cm}~~~~~~~~~~~~~~T > 33000~{\rm K}. \\
\end{cases}
\end{align}

The net heating rate given by Equation~\ref{eq:total_heatcool} enters into the equations of hydrodynamics via the energy equation,
\begin{equation}
\rho\frac{D}{Dt}\left(\frac{e}{\rho}\right)=-P\nabla\cdot v + \rho\mathcal{L},
\end{equation}
where $\rho$ is the mass density, $P$ is the pressure, $v$ is the velocity, $e$ 
is the internal energy and $\mathcal{L}$ is the net heating rate with units of $\rm{ergs~s^{-1}~g^{-1}}$. The plasma is treated as an ideal monatomic gas, with $\gamma=5/3$. 

\subsection{Boundary conditions and the computational grid}
\label{sec:BCs}
The computational grid consists of one quadrant of a circle, with logarithmically spaced cells in the $r$ and $\theta$ direction. A reflecting boundary condition (BC) is used along the mid-plane and an 
axisymmetric BC along the axis of rotational symmetry. Thus we are simulating a 2.5D spherical region  centred on the accretor. The radial grid is 
finest towards the centre
of the grid, with $dr_{k+1}/dr_{k}=1.02$. The polar grid runs from 
$\rm{0\degree}$ to $\rm{90\degree}$ in polar angle, with a spacing 
$d\theta_{k+1}/d\theta_{k}=0.95$. This gives more resolution towards the mid-plane, 
where we expect the density, velocity and temperature to change most quickly.
In addition, we adopt outflow BCs at the inner and outer radial edges, and a minimum density of $10^{-22}\rm{g~cm^{-3}}$ is imposed throughout the grid. In the IRT simulations, we place a non-radiating, reflecting accretion disc at the mid-plane out to a radius of $2R_{IC}$. This represents
a reasonable maximum radius for a stable disk in this type of system (H17).

The density at the mid-plane of the grid out to a radius of $2R_{IC}$ is also fixed and essentially acts as a mass reservoir for the simulation. This density has to be high enough to ensure that the simulation domain encompasses the critical acceleration zone of the outflow, but not so high as to include the entire body of the hydrostatic accretion disc. We satisfy these requirements by setting the mid-plane density on the basis of thermal stability considerations.

\begin{figure}
\includegraphics[width=\columnwidth]{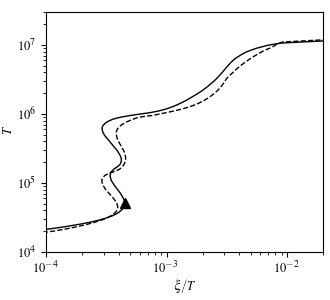}
\caption{Stability plots generated using \textsc{python} (solid line) and
\textsc{cloudy} (dashed line). The symbol highlights the location of the end of the 
lower temperature stable branch for the \textsc{python} simulation.}
\label{figure:stab}
\end{figure}

Figure \ref{figure:stab} shows a set of thermal equilibrium curves generated from two different codes (\textsc{python} and \textsc{cloudy}) for different ways of calculating the heating and cooling rates. Each curve corresponds to an estimate of the
equilibrium temperature as a function of $\xi/T$, i.e. the temperature for which
heating equals cooling. The two codes give slightly different results, which is not unexpected since they use somewhat different atomic data, and the heating and cooling rates are very sensitive to exactly what processes and species are included. 

In both calculations, we see that there is a low temperature branch at the left
hand side, where the equilibrium temperature eventually drops to about 10000~K. 
This gas can
usefully be thought of as being part of the disc atmosphere. There is also
a high temperature branch where the plasma reaches the Compton temperature. In
between, the plasma becomes thermally unstable, sometimes being in a state when 
increasing the temperature actually increases the heating rate, allowing gas
to heat up very rapidly. This is the region of interest for a thermal 
wind, since it is this heating that leads to expansion and therefore acceleration. Moreover, since we are interested in modelling the outflowing gas, rather than the static disc atmosphere, the mid-plane in our simulation would ideally correspond to the upper edge of the hydrostatic disc, i.e. to the hot end of the lower stable branch. 

In the optically thin limit, we can achieve this by computing the critical ionization parameter, $\xi_{cool,max}$, at the end of this branch. We can then set the mid-plane density at all radii to the value necessary to give $\xi =  \xi_{cool,max}$. This is how we set the BC for the mid-plane density in H17. However, in the RHD simulations discussed here, we allow for the attenuation of the radiation field incident on a given grid cell by all of the intervening material along the line of sight to the central source. We also include electron scattering, which can be important in allowing otherwise shadowed parts of the wind to be illuminated indirectly \citep{2010MNRAS.404.1369S,2014ApJ...789...19H}.

As a result of these two effects, the physical properties of the gas and the radiation field are coupled, and it is not possible to calculate $\xi_{cool,max}$ and the corresponding mid-plane density a priori. We can nevertheless safely adopt the mid-plane density estimated for optically thin conditions, since attenuation by intervening material can only {\em reduce} the strength of the incident radiation field and the associated ionization parameter. Setting the mid-plane density in this way therefore guarantees that $\xi < \xi_{cool,max}$. This is at the expense of including at least some material from further down on the lower stable branch in the simulation, i.e. material that belongs to the hydrostatic disc.

For the heating and cooling rates calculated with \textsc{python} in Figure~\ref{figure:stab}, we find $\xi_{cool,max}=22.7$ in the optically thin limit, with an equilibrium temperature $T_{cool,max}=50000$~K (marked with a triangle in Figure~\ref{figure:stab}). The corresponding local mid-plane density is then calculated via
 \begin{equation}
\rho(r)=\rho_0\left(\frac{r}{R_{IC}}\right)^{-2},
\label{equation:midplane_rho}
\end{equation}
with $\rho_0=1.6\times10^{-11}~\rm{g\, cm^{-3}}$. Here, $R_{IC}$ is the Compton radius, i.e. the radius at which the local isothermal sound speed (at the
Compton temperature, $\rm{T_{IC}}$) is equal to the escape velocity,
\begin{equation}
R_{IC} = \frac{GM_{BH}\mu m_H}{k_BT_{IC}}.
\end{equation}

\subsection{Summary of the simulations}

All of the key parameters that define our simulations and describe the resulting outflows are summarized in Table~\ref{table:wind_param}. Three different runs are listed in Table~\ref{table:wind_param}: (i) HD+\textsc{cloudy}$_{\tau\!\ll\!1}$ -- a pure hydrodynamic simulation in which radiative heating and cooling rates were calculated in the optically thin limit using \textsc{cloudy}; (ii) HD+\textsc{python}$_{\tau\!\ll\!1}$ -- a pure hydrodynamic simulation in which radiative heating and cooling rates were calculated in the optically thin limit using \textsc{python}; (iii) RHD -- our new radiation-hydrodynamic simulation in which radiative heating and cooling rates were calculated accounting for attenuation (i.e. finite optical depth) in the wind using \textsc{python}. 

The HD+\textsc{cloudy}$_{\tau\!\ll\!1}$ and HD+\textsc{python}$_{\tau\!\ll\!1}$ simulations are included as benchmarks for assessing the importance of radiative transfer effects in the RHD simulation. The HD+\textsc{cloudy}$_{\tau\!\ll\!1}$ model is identical to the simulation shown in 
H17. The HD+\textsc{python}$_{\tau\!\ll\!1}$ calculation serves to confirm that, in the optically thin limit, we obtain very similar results when we use \textsc{python} instead of \textsc{cloudy}.  Since the HD+\textsc{cloudy}$_{\tau\!\ll\!1}$ and HD+\textsc{python}$_{\tau\!\ll\!1}$ runs are so similar, we will only plot and discuss the HD+\textsc{python}$_{\tau\!\ll\!1}$ version below. 

The third simulation in Table~\ref{table:wind_param} is our new RHD calculation, where we iterate between HD and IRT steps. This allows us to relax the $\tau \ll 1$ assumption and to account for the attenuation of radiation field as it passes through the outflow.

\subsection{Limitation of the simulation: neglect of wind radiation.}

Perhaps the most significant approximation we still have to make is that the material in our simulation is optically thin {\em to its own radiation}. Thus while we calculate the full, frequency dependent spectrum produced by the gas in each grid cell self-consistently, we only use this to estimate the local radiative cooling. We do not consider the effect of cooling radiation produced in one grid cell on the thermal and ionization state of material in other grid cells. 

This approximation is necessary because cells near the mid-plane in our simulations tend to become highly optically thick, a regime that {\sc python} and most other Monte-Carlo based radiative transfer do not (yet) handle very well. The underlying problem is that, in order to capture the acceleration zone of the outflow, our simulation has to include at least some of the hydrostatic disc atmosphere. It is these deep-lying regions that cause numerical problems. 

In order to check whether this approximation affects our results, we have carried out test calculations with {\sc python}. Starting from a converged snapshot model, we excise the troublesome $\tau >> 1$ disc regions, but then add a thin $T_{eff} \simeq 10^4$~K blackbody disc as a photon source "by hand". We find that the heating and cooling rates in the outflow (including the most important acceleration zone) following the addition of this extra source of photons do not change significantly. We therefore expect that the wind's own radiation field would not significantly affect the overall outflow properties. This expectation is supported by results from earlier 1D 
simulations which also
showed that the inclusion of disk radiation did not
significantly alter the structure or temperature
balance of less dense material overlying it \citep{1994ApJ...431..273K}.

\subsection{Comparing to observations: predicted absorption lines}

We generate synthetic absorption lines from our RHD simulation using a ray-tracing method similar to that described in H17. Briefly, we use a converged snapshot of the simulation to provide us with the  density, velocity and ionization state throughout the outflow. For any given ground state transition, we can then calculate the optical depth as a function of velocity, $\tau({v})$, along any given sightline to the centre of the system. In doing so, we also account for the random thermal motions in the outflowing gas by adopting a Doppler-broadened Gaussian as the local line profile. The overall absorption line profile is then estimated as $e^{-\tau(v)}$. 

The inclination towards GRO~J1655-40 has been estimated to be $i =70.2 ^\circ \pm 1.9^\circ$ \citep{2001ApJ...554.1290G}. However, as explained above, our simulations are designed to capture only the uppermost parts of the hydrostatic disc. If the disc itself is flared with a characteristic half-opening angle of $\gamma$, it would be more correct to adopt $i = 70^\circ + \gamma$ in the spectral synthesis calculations. We will therefore show results for both $i = 70^\circ$ and $i = 80^\circ$ when comparing model spectra against observations. 

\section{Results}
\label{section:results}

\begin{figure*}
\includegraphics[width=\columnwidth]{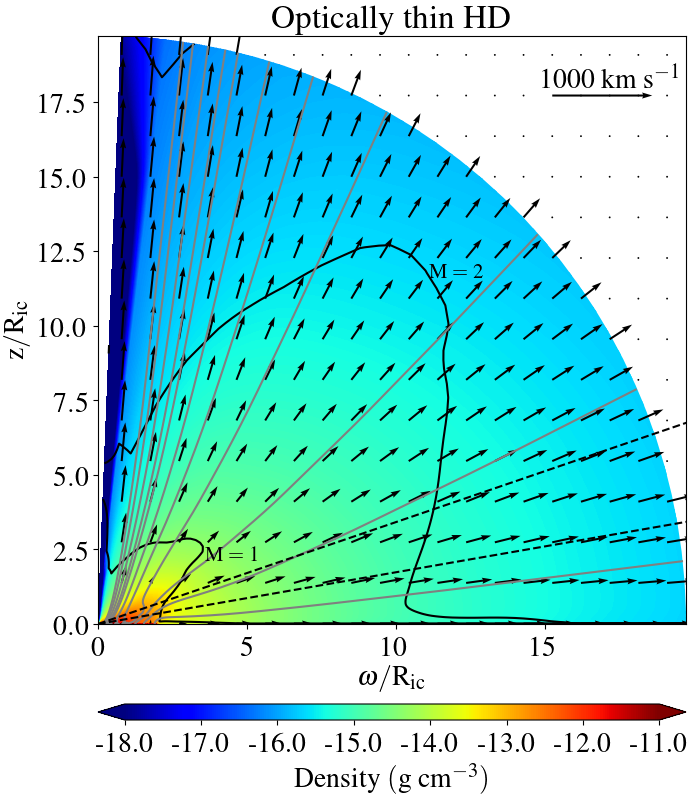}
\includegraphics[width=\columnwidth]{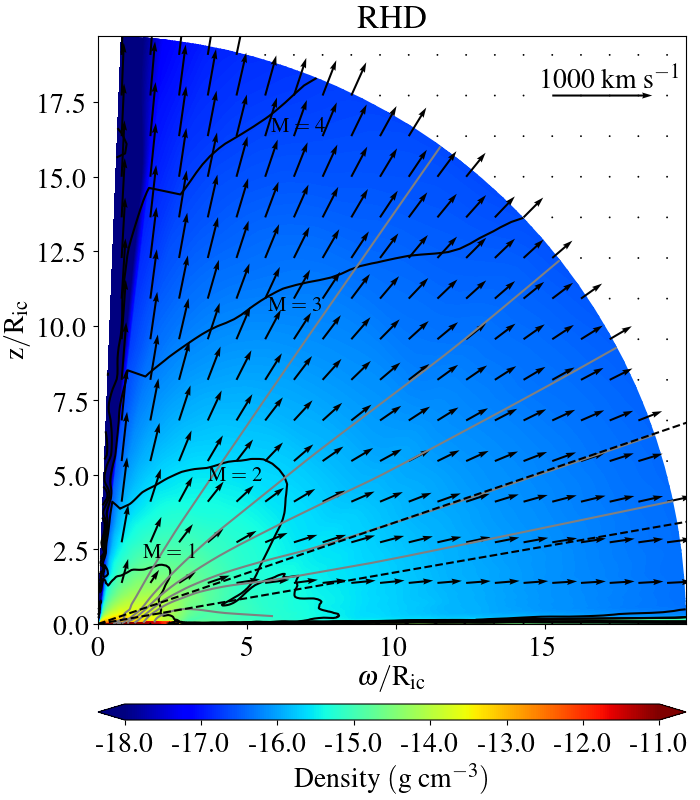}
\includegraphics[width=\columnwidth]{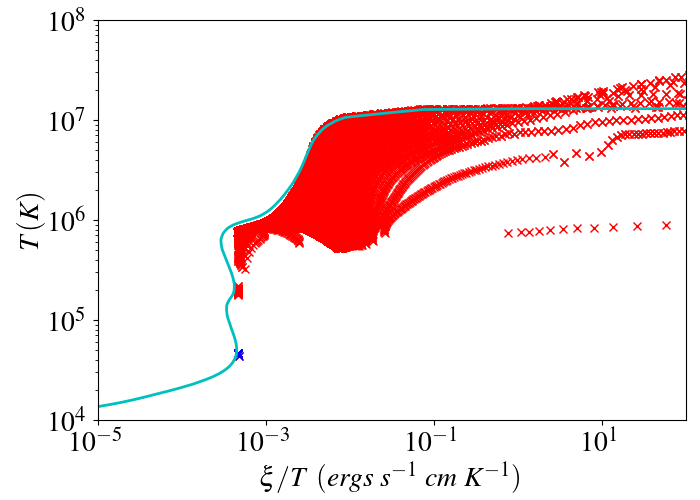}
\includegraphics[width=\columnwidth]{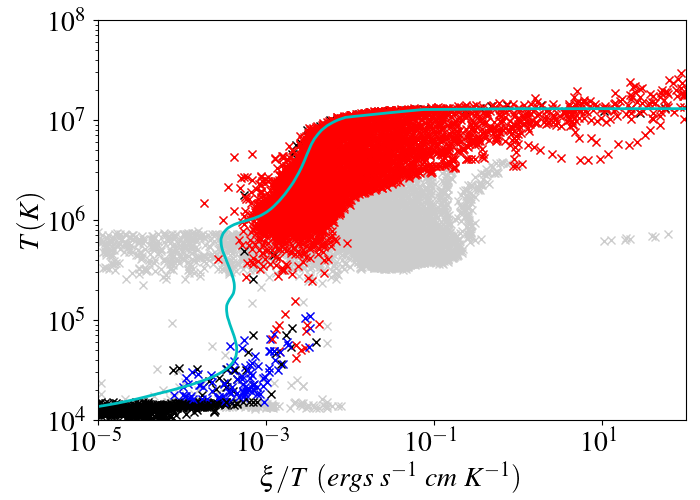}

\caption{Upper panels: The density and velocity structure of the stable final state of 
the simulation. Grey lines show streamlines, and the black line shows the 
location of the 
Mach surfaces. The two 
dotted
lines show the location of the $70\degree$ and $80\degree$ sightlines.
\newline
Lower panels: Stability plots for the same two snapshots. Blue symbols
represents cells at the surface from which the wind is launched. Red
symbols are for cells in the wind and black symbols are for cells below
the wind-launch surface. The grey points in the right lower panel are
for points outside the radial extent of the disc but at high inclination
angle and so still shadowed by it.}
\label{figure:wind_small_image}
\end{figure*}

Figure \ref{figure:wind_small_image} shows a side-by-side comparison of the disc wind properties in a pure HD simulation (HD+\textsc{python}$_{\tau\!\ll\!1}$) versus those in our new RHD simulation. The upper panels of Figure \ref{figure:wind_small_image} show the final (steady-state) density and velocity structure, while the lower panels show the corresponding thermal stability plots. Note that for the RHD simulation, the ionization parameter $\xi$ is calculated directly from the \textsc{python} IRT simulation, i.e. it is computed from the actual energy packets passing through each cell. Each point in these stability plots represents an individual grid cell, and the solid curves show the optically thin, \textsc{python}-based thermal equilibrium curve. As noted above, the key parameters of the outflows are summarized in Table~\ref{table:wind_param}.

A key feature in the new RHD simulation is a wedge-shaped, high-density region near the mid-plane (which extends out to $r=2R_{IC}$ - the extent of our `mass reservoir' BC). As discussed in Section~\ref{sec:BCs}, the mid-plane density we adopt guarantees that the RHD simulation grid covers the acceleration zone of the outflow, but it also means that this grid must include part of the hydrostatic disc. The formation of a dense, quasi-static region near the mid-plane in the RHD calculation is therefore inevitable. This 
structure broadly mirrors the 'disk' structure seen in 1D simulations \citep{1993ApJ...412..267R,1994ApJ...431..273K,2002ApJ...581.1297J}.

The wedge-like shape of this region can also be understood. X-rays emitted from the central source can only heat the atmosphere until the integrated optical depth along their path reaches $\tau \simeq 1$. For any given photon launch angle with respect to the mid-plane, there is thus a maximum radius out to which these photons can contribute to driving the outflow. Larger disc radii can only be reached by photons emitted with larger launch angles. The aspect ratio corresponding to the wind acceleration zone, $H/R$, must therefore {\em increase} with $R$, i.e. the shape of this zone must be (slightly) convex. At the outer edge of the 
hydrostatic disk ($r\simeq 10^{12}$cm), the disk has an opening angle of about 4\degree. This is reasonably close to 
the disk opening angle of about 2.3\degree~ inferred for the LMXB
Cyg X-2 \citep{1990A&A...235..162V}

Thermally stable grid cells within the body of the disc are represented by black symbols on the stability plot (bottom panels of Figure~\ref{figure:wind_small_image}). They all have a temperature of around 12000~K, with the radiative heating coming from the Auger ionization of metals. This process is associated with hard X-ray photons that are able to penetrate deep into the disc. The vertical density profile in the dense hydrostatic region is fairly flat, with cells maintaining roughly the same density as the mid-plane cell below. The boundary of the hydrostatic disc in the RHD simulation is very sharp, a 
phenomenon also seen in 1D simulations 
\citep{1993ApJ...412..267R,1994ApJ...431..273K,2002ApJ...581.1297J}. In
those simulations, the transition between the 
cool 'disk' and the hotter coronal gas above it 
was unresolved. Unsurprisingly, we also do 
not resolve the transition, however changing the 
resolution at the transition region by increasing
the number of cells in the $\theta$ direction has no
effect on the overall structure of the resulting
wind. 

Above the disk, the temperature rapidly rises
to a few million K. It is clear from the lower panels of Figure \ref{figure:wind_small_image} that relatively few cells actually reach the stability curve, where radiative heating and cooling are in balance. This is because for the 
bulk of the simulation domain, adiabatic cooling is balancing radiative heating. Referring back to the 1D simulations of accretion disks, this means that
most of the gas in the domain can be best described as an expanding disk atmosphere. 
This is exactly as one would expect, since we are concentrating on the outer parts of
the disk where those simulations showed that, even neglecting adiabatic cooling, one only reaches the Compton temperature at a distance roughly equal to the launching radius. Another interesting
observation is that the throughout the simulation domain, X-ray heating and Bremsstrahlung cooling are important contributors to the balance; there are few locations where the temperature is set only by Compton effects

The density in the wind above this boundary is lower than in the optically thin HD run. This reduced density is reflected in a significantly ($\gtrsim 5\times$) reduced mass-loss rate of the RHD model compared to the pure HD simulations. However, even this reduced mass-loss rate amounts to more than twice the accretion rate onto the central object (as implied by our adopted luminosity). This `wind efficiency' of about 2 is very close to that predicted by \cite{2018MNRAS.473..838D} for our disk size
($2R_{IC}$) and our luminosity ($0.5L_{crit}$, where $L_{crit}\sim 0.03T_{IC,8}^{-0.5}L_{Edd}$).

The terminal outflow velocity in the RHD model is comparable to that in the optically thin HD simulations, reaching $v_\infty \simeq 300~{\rm km~s^{-1}}$. This is in line with the expectation that $v_\infty \simeq v_{th}$, since the characteristic temperature in outflowing material is $T \simeq 5\times 10^6$~K (see Figure~\ref{figure:wind_small_image}). At this temperature, the rms thermal speed of atoms is $v_{th} = \sqrt{3k_BT/\mu m_p} \simeq 450~{\rm km~s^{-1}}$, where $\mu \simeq 0.6$ is the mean molecular weight. 

The dashed lines in the upper panels of Figure~\ref{figure:wind_small_image} show direction vectors corresponding to high-inclination lines of sight with $i = 70\degree$ and $80\degree$. They are provided here to highlight the inclinations for which we will show synthetic absorption line profiles in Section~\ref{section:line_prof}. These inclinations are typical of the soft-state LMXBs in which wind-formed absorption features have been observed \citep{2012MNRAS.422L..11P}. 

\begin{table}
\begin{tabular}{p{3.0cm}p{0.9cm}p{1.0cm}p{0.9cm}}
\hline 
Simulation Method & HD & HD & RHD \\ 
\hline 
\pbox{20cm}{Heating/Cooling \\ Rates} & \pbox{20cm}{\textsc{cloudy} \\ $\tau \ll 1$} & \pbox{20cm}{\textsc{python} \\ $\tau \ll 1$} & \pbox{20cm}{\textsc{python} \\ any~$\tau$} \\ 
\hline 
\hline Physical Parameters & & & \\ \hline
$\rm{M_{BH}~(M_{\odot})}$  & 7 & 7 &7\\
$\rm{T_x~(10^7~}$K)  & 5.6 & 5.6  & 5.6\\
$\rm{L_x~(10^{37}}~\rm{ergs~s^{-1}})$&  3.3& 3.3 &3.3\\
$\rm{\log(\xi_{cold,max})}$&   1.226 & 1.35 & -- \\
$\rm{T_{eq}(\xi_{cold,max})~(10^3~K)}$ &  43.7 & 50.7 & 50.7 \\
$\rm{\rho_0~(10^{-12}~g~cm^{-3})}$  & 20.3 & 16.0 & 16.0 \\
$\rm{R_{IC}~(10^{11}~cm)}$  & 4.82 & 4.82 & 4.82 \\
\hline
\multicolumn{4}{l}{Grid parameters}\\
\hline
$\rm{R_{min} (10^{10}~}$cm) & 2.4 & 2.4 &2.4 \\
$\rm{R_{max} (10^{10}}~$cm) &  960& 960 & 960\\
$\rm{R_{disc} (10^{10}}~$cm) & 96& 96 & 96\\
$\rm{R_{ratio}}$ &1.05&1.05&1.05\\
$\rm{N_R}$  & 200& 200&200\\
$\rm{\theta_{min}}$ & 0.0& 0.0&0.0\\ 
$\rm{\theta_{max}}$ & 90.0& 90.0&90.0\\
$\rm{\theta_{ratio}}$ & 0.95& 0.95 & 0.95\\
$\rm{N_{\theta}}$ & 100& 100 & 100\\
\hline
\multicolumn{4}{l}{Derived wind properties}\\
\hline 
\multicolumn{4}{l}{$\rm{V_r~(\rho>1\times10^{12})}$}  \\
$\rm{max,blueshifted~({km~s^{-1}})}$ &  19 & 30  & 47  \\
$\rm{V_r(max,\theta>60\degree~({km~s^{-1}})}$ &  290   & 280    & 260 \\
$\rm{N_H~(70\degree)~(10^{22}}~\rm{cm^{-2})}$ & 13 & 12 & 2.0 \\
$\rm{N_H~(80\degree)~(10^{22}}~\rm{cm^{-2})}$  & 63 & 55 & 4.4 \\
${\dot{\rm{M}}_{\rm{wind,outer}}~(10^{18}~g~s^{-1}})$  &  6.7 & 5.6  & 1.1 \\
${\dot{\rm{M}}_{\rm{wind,outer}}~(\dot{\rm{M}}_{\rm{acc}})}$  & 15 & 13 & 2.5  \\
${0.5\dot{\rm{M}}\rm{V_r^2}~(10^{32}}~\rm{erg~s^{-1}})$  &  24 &  21 & 4.2  \\

\hline
\end{tabular}
\caption{Parameters adopted in the simulations, along with key properties of the resulting outflows.}
\label{table:wind_param}
\end{table}

\section{Discussion}
\label{section:discussion}

\subsection{The impact of attenuation}

As we have seen, our RHD simulation produce a disc wind with a roughly $5\times$ lower mass-loss rate compared to optically thin, pure HD calculations. Thus while the pure HD calculations are useful for gaining qualitative insights into the nature of thermally-driven winds, radiative transfer effects are not negligible and must be taken into account in quantitative modelling efforts. 

\begin{figure}
\includegraphics[width=\columnwidth]{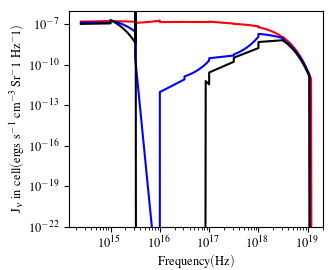}
\caption{Mean intensity in a set of cells at $\omega=1.5R_{IC}$. The blue curve 
is for a cell at the upper surface of the dense disc at the mid-plane. The red curve is for 
the lower density cell directly above it, and the black curve is for the disc cell directly 
below it.}
\label{figure:cell_spec}
\end{figure}

In order to illustrate {\em how} optical depth effects affect the properties of the resulting outflow, Figure \ref{figure:cell_spec} shows the radiation field seen in a set of typical cells in our RHD simulation. These curves show the mean intensity models computed by \textsc{python} for a set of cells near the acceleration zone of the outflow, i.e. straddling the upper surface of the disc-like structure towards the mid-plane.

These mean intensity models are calculated during the \textsc{python} cycles by logging each Monte Carlo energy packet which passes through (or is absorbed by) each cell in the model. These packets are initially generated over a wide continuous range of frequencies, but in individual cells the distribution is modelled in a series of frequency bins (two per frequency decade) using either a power law or exponential model. Deep absorption features are captured by truncating the frequency bins on the fly to take account of any frequencies where no packets are seen at all. This technique captures the gross features of the mean intensity very flexibly, and permits the calculation of photo-ionization rates for complex radiation fields reasonably accurately.

The blue curve in Figure~\ref{figure:cell_spec} corresponds to a cell at the upper surface of the  dense disc, while the red curve corresponds to the cell directly above it. The optical depth from the central object to this latter cell is small, so the red curve essentially shows the expected radiation field at this location in the absence of attenuation. This comparison clearly shows that the radiation field reaching the acceleration zone of the outflow is significantly attenuated, primarily due to photo-ionization of hydrogen. The result is a lower X-ray heating rate near the disc surface, and this is likely to be the main reason for the lower mass-loss rate seen in the RHD simulation. 

The black curve in Figure \ref{figure:cell_spec} is typical of the radiation field in cells lying deeper within the disc. At these depths, there are essentially no hydrogen ionizing photons left,  so this dense part of disc is relatively cool and mostly neutral in the RHD simulation. The dominant  ionization and heating process in these cells is due to inner shell photo-ionization of metals by X-ray photons. This provides enough heating to keep the temperature at around 12,000K - the temperature of the cool stable branch in the bottom right panel of Figure~\ref{figure:wind_small_image}.

Figure~\ref{figure:cell_spec} shows that the radiation field in the disc/wind transition region can be {\em strongly} modified by absorption along the line of sight. This in turn reduces the ionization rate in this region, and hence reduces the local heating and cooling rates (predominantly photo-ionization heating and recombination/line cooling). In light of Figure~\ref{figure:cell_spec}, the significant differences in the physical outflow properties seen in our HD and RHD simulations should not come as a surprise.

Previous simulations of disk-winds have demonstrated that indirect illumination by scattered radiation can also be important \citep{2010MNRAS.404.1369S,2014ApJ...789...19H}. In  our RHD simulation, the ionizing photon field in cells in the disc, with its high column densities towards the central source, is indeed dominated by radiation scattered from electrons in the wind back down towards the disc. At the boundary between the disc and the wind, this effect is smaller, with at most 30\% of the ionizing photons in these cells arriving indirectly. Indirect irradiation is therefore a secondary effect in this particular simulation, but the ability to include it in our RHD calculations is an important advance.

\subsection{Predicted line profiles}
\label{section:line_prof}

\begin{figure}
\includegraphics[width=\columnwidth]
{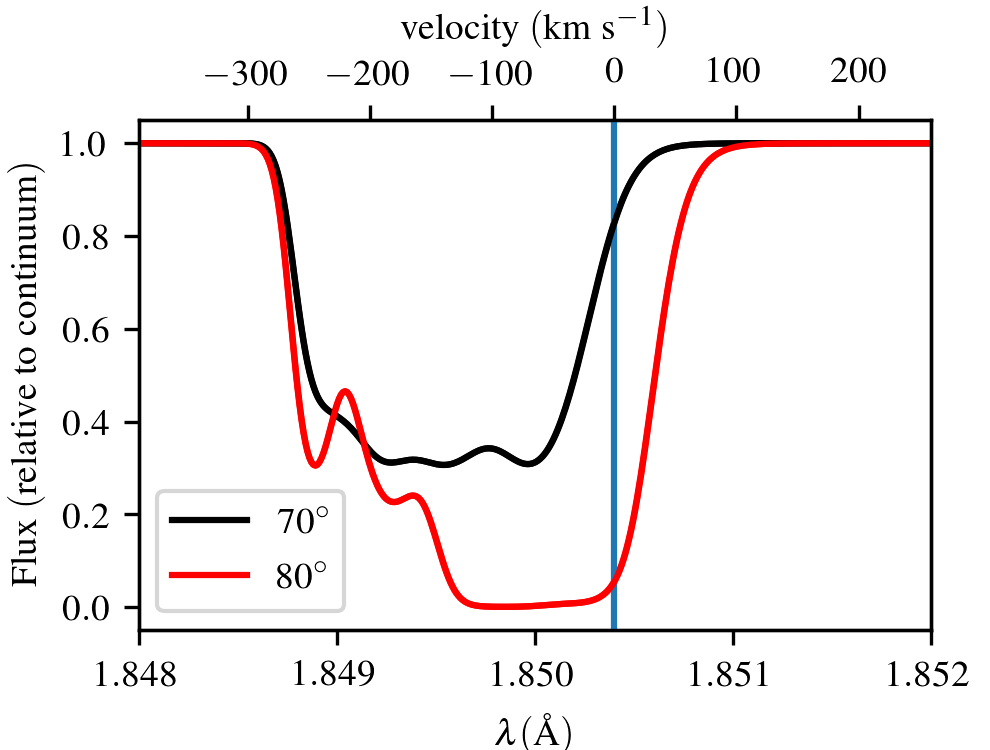}
\caption{Predicted absorption line profiles for the Fe~\textsc{xxv} Lyman~$\rm{\alpha}$ transition at 1.8504~{\AA}, as viewed from $i = 70\degree$ and $i = 80\degree$.}
\label{figure:spectrum_zoom}
\end{figure}

Figure~\ref{figure:spectrum_zoom} shows the intrinsic profile of the wind-formed Fe~{\sc xxv} 1.85~\AA\ absorption line in our simulation, i.e. before taking any instrumental resolution into account. This shows that there is significant optical depth out to roughly $v_\infty \simeq 300~{\rm km~s^{-1}}$. The positive velocity tail of the profile -- reaching $\simeq +100~{\rm km~s^{-1}}$ -- is due to the significant thermal Doppler broadening in the outflowing gas. 

\begin{figure}
\includegraphics[width=\columnwidth]
{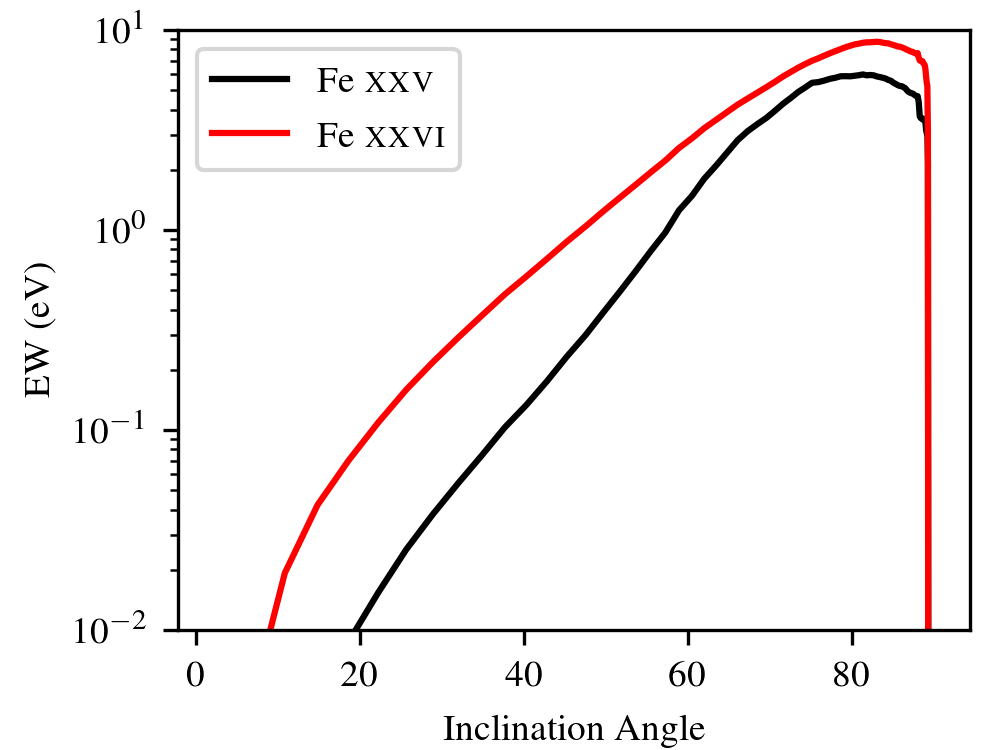}
\caption{Absorption line equivalent width associated with Fe~\textsc{xxv} 1.85~{\AA} ~and the Fe~\textsc{xxvi} 1.78~\AA\ doublet.}
\label{figure:EW}
\end{figure}

Figure~\ref{figure:spectrum_zoom} also illustrates that line {\em strength} is a clear function of inclination, with higher-inclination sightlines producing stronger lines than lower-inclination ones. 
This trend is a natural consequence of the equatorial nature of the outflow, since higher (column) densities are preferentially found near the disc plane (see Figure~\ref{figure:wind_small_image}). We illustrate this more directly in Figure~\ref{figure:EW}, which shows the dependence of the predicted Fe~{\sc xxv}~1.85~\AA\ and Fe~{\sc xxvi} 1.78~\AA\ equivalent widths on inclination. Observable features with ${\rm EW} \gtrsim 1~{\rm eV}$ are only predicted for inclinations $i \gtrsim 60\degree$. This is comparable to the range of inclinations for which wind-formed absorption lines have been observed (Ponti et al. 2012).

\subsection{Comparison to Observations}
\label{section:comp_to_obs}

As discussed in Section~\ref{section:GRO_params}, the parameters adopted in our RHD simulations roughly match those of GRO~J1655-40 in its soft-intermediate state. Are the Fe~{\sc xxv} and Fe~{\sc xxvi} absorption lines we predict consistent with the {\sc Chandra} HETG observations obtained during this state?

\begin{figure}
\includegraphics[width=\columnwidth]{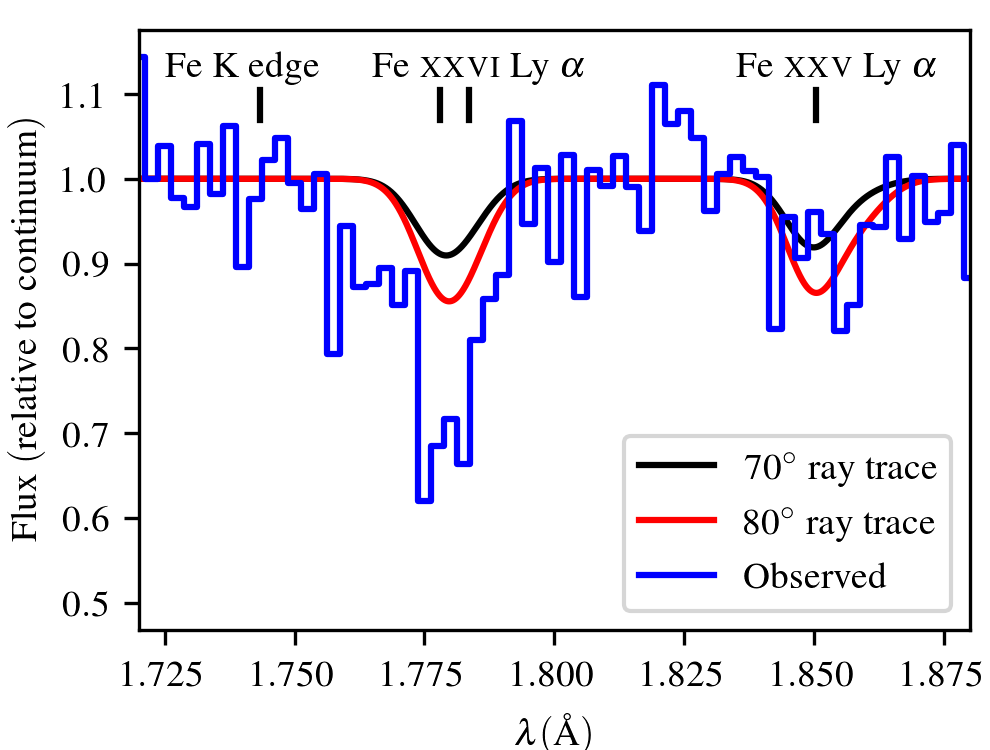}
\caption{$80\degree$ `ray-traced' spectrum around the ground state Fe  ~\textsc{xxv}  and ~\textsc{xxvi} resonance lines (grey line). The red line is the effect
of convolving this spectrum with a 0.012~{\AA} FWHM Gaussian to represent the 
resolution of the HETG and the blue line is the March 2006 \emph{Chandra}
observation of GRO J1655-40. This is scaled to a continuum based
on the flux at 1.7~\AA\ and 1.9~\AA\ to allow comparison with the ray-traced
spectrum.}
\label{figure:spectrum}
\end{figure}

Figure~\ref{figure:spectrum} shows a direct comparison between predicted and observation absorption line profiles. Here, we have convolved the predicted profiles with the {\sc Chandra} HEG resolution of 0.012~\AA (FWHM), corresponding to $\Delta v \simeq 2000~{\rm km~s^{-1}}$ at 1.8~\AA. Even though the lines are weak, the data are noisy, and the profiles are unresolved,
it is encouraging that the simulations predict line strengths and line ratios that are within a factor of two or so of the observed ones. In assessing the quality of this agreement, it is important to remember that Figure~\ref{figure:spectrum} is not a {\em fit} to the observations. No parameters were adjusted to optimize the match to the data. Given all of the uncertainties and approximations associated with such a comparison, this level of agreement is promising.

\cite{2002ApJ...567.1102L} and \cite{2012MNRAS.422L..11P} have used the wind-formed Fe~{\sc xxv} and Fe~{\sc xxvi} absorption lines in LMXBs to estimate the mass-loss rates in these outflows. Using our synthetic line profiles, we can test how well this method recovers the true mass-loss rate in the RHD simulation. Briefly, the method is based on combining the mass continuity condition, 
\begin{equation}
\dot{M}_{wind} = \Omega R^2 n m_p v_{wind}, 
\end{equation}
with the definition of the ionization parameter, $\xi$, given in Equation~\ref{eq:ip}. The result is
\begin{equation}
\dot{M}_{wind} = \Omega m_p v_{wind} L_x / \xi.
\end{equation}
Here, $\Omega$ is the solid angle subtended by the outflow, and $m_p$ is the proton mass. 

Observationally, $L_x$ is determined from the X-ray continuum and $v_{wind}$ from the intrinsic width and/or blueshift of the absorption line. The solid angle of an equatorial outflow along the disc plane with half-opening angle $\alpha$ is just $\Omega = 4\pi \sin{\alpha}$. As noted above, the range of inclinations for which wind-formed absorption lines have been observed suggests $\alpha \simeq 30\degree$ \citep{2012MNRAS.422L..11P}, in line with what we find in our RHD simulation. The solid angle subtended by the outflow is therefore roughly $\Omega \simeq 2\pi$.

\begin{figure}
\includegraphics[width=\columnwidth]{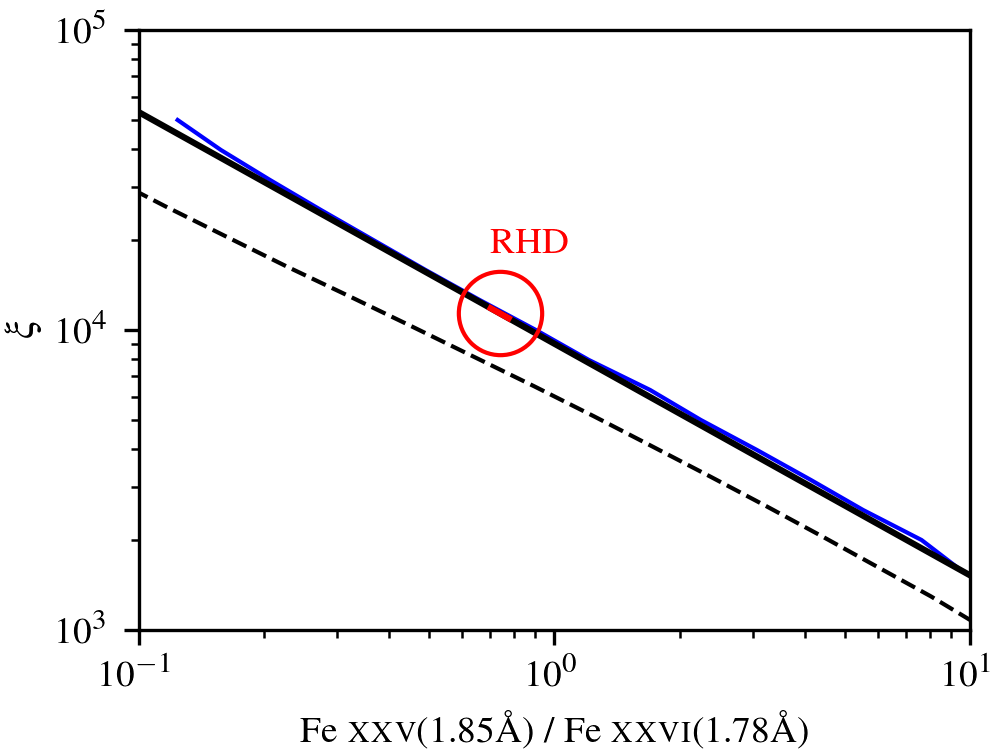}
\caption{Relationship between ratio of Fe\textsc{xxv} and Fe\textsc{xxvi} Lyman
$\rm{\alpha}$ equivalent widths and ionization parameter - based on data from
\citet{2001ApJS..133..221K}. Dashed line shows relationship based on data from
\textsc{python} simulations using the same bremsstrahlung SED as used in the 
RHD simulation}
\label{figure:line_rat_vs_IP}
\end{figure}

This leaves only the ionization parameter, $\xi$, to be estimated. This is done by measuring the Fe~{\sc xxv} / Fe~{\sc xxvi} line ratio and comparing it to single-cloud, optically thin ionization calculations for a particular reference SED \citep[e.g.][]{2001ApJS..133..221K}. The dependence of ionization parameter on the Fe line ratio in the calculations of \citet{2001ApJS..133..221K} is shown by the solid line on Figure~\ref{figure:line_rat_vs_IP}. For the line ratios we obtain from our synthetic spectra for sightlines in the range $70\degree \leq i \leq 80\degree$ (about 0.7-0.8), we would estimate $\xi \simeq 1.1 \times 10^4$ for the line-forming region in our outflow. 

As noted by \citet{2012MNRAS.422L..11P}, the actual SED does, of course, have a significant impact on the line ratios. For reference, we therefore also show in Figure~\ref{figure:line_rat_vs_IP} the relationship between the Fe line ratio and ionization parameter as calculated with \textsc{python} for the actual (unattenuated) SED used in our RHD simulation. This relationship would give an estimate of $\xi \simeq 7 \times 10^3$.

Adopting $L_x = 3.3 \times 10^{37}~{\rm erg~s^{-1}}$, $v_{wind} \simeq 300~{\rm km~s^{-1}}$, $\Omega \simeq 2\pi$ and $\xi \simeq 10^{4}$ as characteristic numbers for the wind in the RHD simulation, we obtain an ``empirical'' estimate of $\dot{M}_{wind} \simeq 10^{18}~{\rm g~s^{-1}}$. This is reassuringly close to the actual mass-loss rate through the boundary of our computational grid,  $\dot{M}_{wind} = 1.1 \times 10^{18}~{\rm g~s^{-1}}$, and demonstrates the robustness of this method of measuring mass-loss rates.

\begin{figure*}
\includegraphics[width=0.8\textwidth,angle=270]{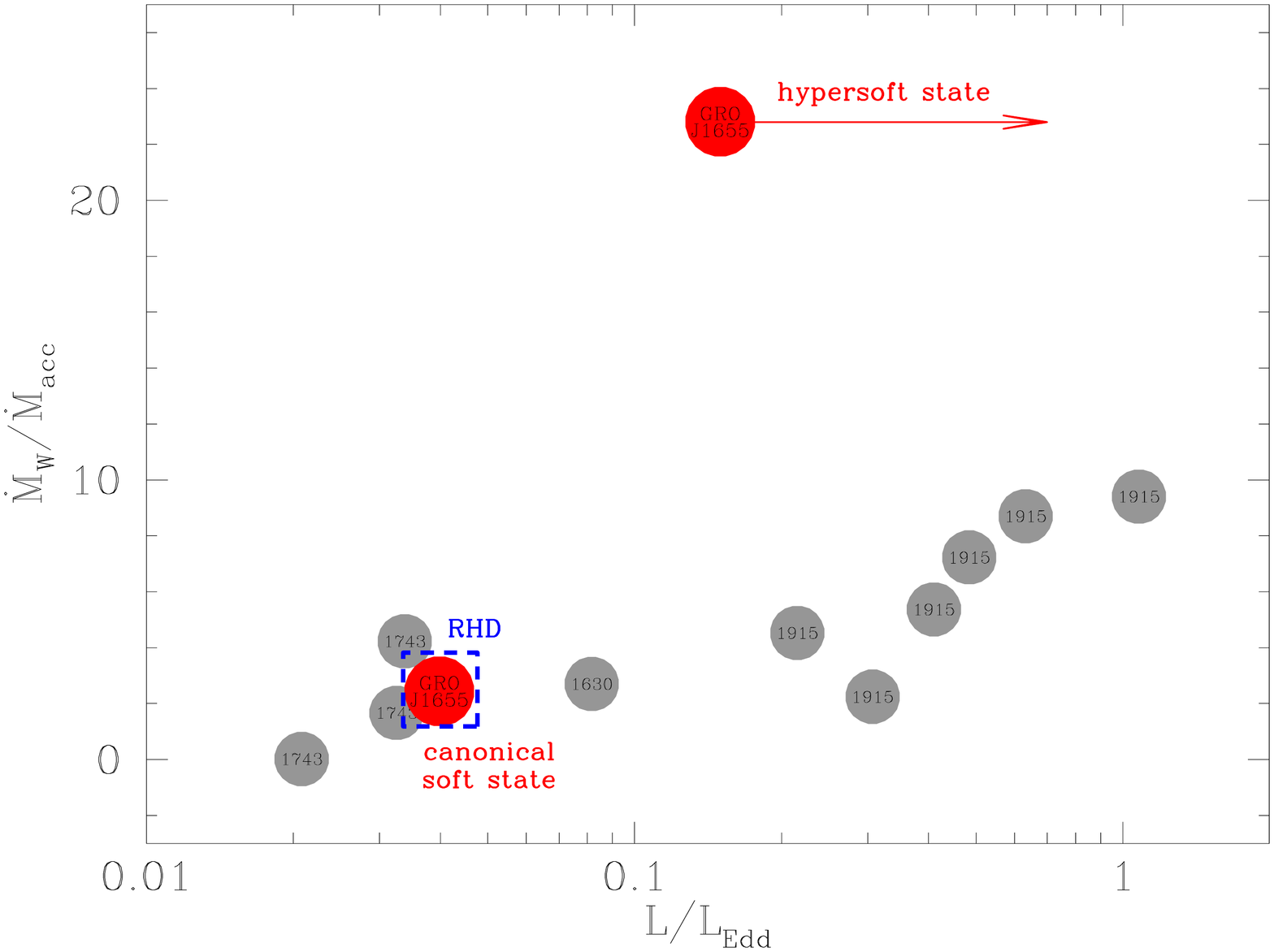}
\caption{$\dot{M}_{wind}/\dot{M}_{acc}$ vs. luminosity based upon figure 5 in \citet{2012MNRAS.422L..11P}. Filled circles are empirical values obtained from {\em Chandra} HETG data for several LMXBs. The labels inside the circles refer to the specific object each measurement refers to (1743 = H1743-322; 1630 = 4U~1640-47; 1915 = GRS~1915+105; GRO~J1655 = GRO~J1655-40). The two red circles represent the measurements obtained for GRO~J1655-40 in the hypersoft and soft-intermediate states. As discussed in Section~\ref{section:GRO_params}, the measurement for the hypersoft state might be a lower limit (if the outflow in this state was Compton-thick). The green square shows the ratio of mass-loss rate to accretion rate predicted by an optically thin, pure HD simulation. The blue square shows the same ratio for our RHD simulation, demonstrating the excellent agreement with the empirical result for GRO~J1655-40 in the soft-intermediate state.}
\label{figure:ponti}
\end{figure*}

With this in mind, how close is the mass-loss rate in our model to the observationally inferred one for the soft-intermediate state? \citet{2012MNRAS.422L..11P} estimated luminosities and mass-loss rates from a set of {\em Chandra} HETG observations of black-hole LMXBs, including GRO~J1655-40. We reproduce their results in Figure~\ref{figure:ponti}, highlighting the two measurements they obtained for GRO~J1655-40. These correspond to the two {\em Chandra} HEG spectra shown in Figure~\ref{figure:states}, i.e. they represent the March 2005 soft-intermediate state and the April 2005 hypersoft state. Remarkably, the mass-loss rate we obtain from our RHD simulation is a close match to the rate inferred for the soft-intermediate state of GRO~J1655-40.

It should be noted that at 80\degree, the lines are saturated. In this
case the velocity spread of material contributing to the absorption is more important to the line EW than the mass-loss rate. In addition, the 
Fe \textsc{XXV} inter-combinational line at 1.8595 \AA ~will be blended
with the resonance line. This  line is normally very weak in comparison but if the resonance line is saturated, it could measurably change the EW of the blend. These two effects could slightly alter the line ratios and hence massloss rate.

\section{Summary }
High inclination LMXBs show clear evidence of disc winds when they are in disc-dominated soft states. In order to test whether these winds may be thermally driven, we have developed a new operator-splitting method for carrying out radiation-hydrodynamic simulations of these outflows, based on combining elements of the hydro-code {\sc ZEUS} and the radiative transfer code {\sc PYTHON}. The resulting RHD code allows for arbitrary SEDs and accounts for full frequency-dependent radiation transfer and ionization calculations.

We then carried out RHD simulations of a thermally-driven wind in an LMXB, using the same basic system parameters ($M_{BH}$, L$_X$, SED, etc.) as used in earlier simulations that assumed that outflow to be optically thin throughout. 

Our main conclusions are as follows:

\begin{itemize}

\item{In terms of its geometry and velocity field, the outflow that develops in our RHD simulation is qualitatively similar to the winds seen in pure HD simulations that adopt the optically thin limit. However, the mass-loss rate is much lower, $\dot{M}_{w} \simeq 1 \times 10^{18}$ g~s$^{-1}$ in the RHD simulations, compared to $\dot{M}_{w} \simeq 6 \times 10^{18}$ g~s$^{-1}$ in the optically thin HD calculation. Nevertheless, this reduced mass-loss rate still represents approximately 2.5$\times$ the accretion rate.}

\item{The reason for this reduced mass-loss rate is strong photo-absorption between the X-ray source and the wind-launching region in the model. This decreases the amount of heating in these cells.  Thus properly incorporating radiative transfer into simulations of thermally-driven disc winds is important.}

\item{The system parameters of our fiducial model broadly match those of the well-known
LMXB system GRO J1655-40. The mass-loss rate in our RHD simulation is similar to that estimated by  \citet{2012MNRAS.422L..11P} for GRO J1655-40 in a soft-intermediate state. Moreover, ray-traced absorption line profiles generated for our simulated outflow show reasonable agreement with a {\em Chandra} spectrum obtained from this source in the  soft-intermediate state.}

\end{itemize}

Overall, our results suggest the that thermal driving remains a viable mechanisms for powering the disc wind in (most) LMXBs.

\section{acknowledgements}
Calculations in this work made use of the Iridis4 Supercomputer at the University of Southampton.
NH and CK  acknowledge support by the Science and Technology Facilities Council grant ST/M001326/1. KSL acknowledges the support of NASA for this work through grant NNG15PP48P to serve as a science adviser to the Astro-H project, JHM is supported by STFC grant ST/N000919/1 and SAS is supported by STFC through grant, ST/P000312/1. Finally we thank the referee John Raymond for a very useful review which suggested many interesting comparisons.

\bibliographystyle{mnras}
\bibliography{bibliography}()
\label{lastpage}

% Don't change these lines
\bsp	% typesetting comment

\end{document}